\documentclass[cameraready]{Interspeech}

\title{Improving Code-Switching ASR with Code-Mixing Guided Synthetic Speech}

\author[affiliation={1,2}, orcid=0009-0001-0884-8888]{Yue Heng}{Yeo}
\author[affiliation={1}]{Haoyang}{Li}
 \author[affiliation={1}, orcid=0000-0002-5718-570X]{Yizhou}{Peng}
\author[affiliation={1}, orcid=0009-0000-8699-3362]{Shreyas}{Gopal}
\author[affiliation={1}, orcid=0000-0002-3998-9229, correspondingauthor]{Hexin}{Liu}
\author[affiliation={3}, orcid=0000-0002-7449-5726]{Leibny Paola}{Garcia-Perera}
\author[affiliation={2}, orcid=0000-0001-6872-5153]{Hardik B.}{Sailor}
\author[affiliation={4}, orcid=0000-0003-3742-7510]{Jeremy H. M.}{Wong}
\author[affiliation={1}, orcid=0000-0001-6257-7399]{Eng Siong}{Chng}

\address{
    $^1$ College of Computing and Data Science, Nanyang Technological University, Singapore \\
    $^2$ Institute for Infocomm Research (I$^2$R), A$^{\star}$STAR, Singapore \\
    $^3$ HLT-COE \& CLSP, Johns Hopkins University, USA \\
    $^4$ Google DeepMind, Singapore
}

\email{yuehengyeo001@e.ntu.edu.sg; hexin.liu@ntu.edu.sg}


\usepackage{comment}
\usepackage{multirow}
\usepackage[utf8]{inputenc}
\usepackage{CJKutf8}

\begin{document}

\maketitle
\begin{abstract}
Code-switch (CS) Automatic Speech Recognition (ASR) remains challenging due to limited availability of high quality CS text-speech pairs for training. Although synthetic data augmentation via Text-to-speech (TTS) has been explored, existing CS TTS approaches primarily optimise reconstruction fidelity and do not explicitly enforce language-boundary consistency, thereby limiting their effectiveness for CS ASR augmentation. This paper proposes a code-mixing guided preference-learning framework that steers synthetic speech generation toward improved code-switching fidelity using the Code Mixing Index (CMI). Experiments on the SEAME Mandarin-English conversational corpus demonstrate that the proposed method enhances the utility of synthetic data for ASR fine-tuning. Specifically, when fine-tuning Whisper Large, the proposed approach reduces Mixed Error Rate (MER) from 12.1\%/17.8\% to 8.9\%/14.2\% on the DevMAN and DevSGE sets, respectively.
\end{abstract}

\section{Introduction}

Code-Switching (CS), the alternation of multiple languages within a single utterance, is a phenomenon in multilingual communities~\cite{dogruoz2021survey, liuxsa, liu23_icassp}. Despite substantial advances in Automatic Speech Recognition (ASR), recognising conversational CS speech remains challenging due to language alternation, cross-lingual phonetic interference, and informal speaking styles~\cite{ugan2024decm, liu2024enhancing}. These challenges result from the scarcity of large-scale, high-quality transcribed CS corpora such as SEAME~\cite{lyu2010seame}, limiting the effectiveness of modern data-intensive ASR systems. As a result, strong end-to-end architectures, including large pretrained models such as Whisper, struggle to generalise robustly to spontaneous conversational CS~\cite{gulati2020conformer,radford2023whisper, yang2025whisper_cs_spt}.

To address data scarcity, data augmentation via text-to-speech (TTS) has become a practical strategy for improving ASR performance in low-resource and domain-mismatched settings~\cite{yang2024bridgingdatagap, zevallos2022tts_aug, ueno2021vq_tts_aug, yuen2023asr}. Multilingual and multi-speaker TTS systems enable scalable generation of labelled speech while introducing speaker and acoustic variability, which has been shown to improve ASR robustness and generalisation~\cite{casanova2023tts,wang2023valle,lin2026stepaudio}. In code-switching contexts, prior studies have explored generating synthetic bilingual speech or phrase-mixed code-switching data to compensate for limited code-switching conversational recordings~\cite{yeo2025apsipa,nguyen2025phrasecs, yan2025csfleurs}. These approaches primarily emphasise acoustic diversity or text coverage, demonstrating that synthetic data can reduce recognition errors when combined with real speech. However, existing augmentation strategies generally assume that all synthesised utterances are equally informative, without explicitly assessing whether the synthesised data's language alternation patterns in conversational code-switching structure is accurate.

Code-Mixing Index (CMI) is a text-based metric that provides a simple and interpretable measure of language mixing balance within a tokenized multilingual sequence \cite{gamback2014cmi}. While CMI has been widely used to analyze code-switching text~\cite{srivastava2021calcs_codemix_metrics,rallabandi2018cs_style_from_acoustics}, its potential as a guiding signal for acoustic modeling remains largely unexplored, primarily due to a fundamental representational mismatch: CMI is defined over discrete symbolic tokens with explicit language labels, whereas speech is a continuous acoustic signal without inherent linguistic segmentation. Therefore, it is not trivial to reliably assign frame-level or segment-level language identities directly from the waveform.


We address this challenge by proposing CMI$_{\text{speech}}$, a novel metric that quantifies language mixing within a speech utterance using pseudo frame-level language labels. Leveraging this metric, we design a TTS-based augmentation framework for CS ASR. Specifically, we first optimize a TTS model through preference learning using Direct Preference Optimization (DPO)~\cite{rafailov2023dpo}, incorporating CMI$_{\text{speech}}$ into preference pair construction to explicitly enforce code-switching consistency. Unlike prior DPO-based TTS methods that focus on perceptual quality or intelligibility without directly constraining cross-lingual boundaries, our approach favors samples exhibiting more realistic language mixing as measured by CMI$_{\text{speech}}$. Subsequently, we fine-tune the downstream ASR using synthetic data generated by the optimized TTS model. By allowing CS linguistic signals to guide acoustic model optimisation, our framework produces synthetic CS speech that better preserves language boundaries and mixing patterns, improving ASR performance on the CS problem.

\section{Preference Learning}

Preference learning has recently emerged as a scalable preference alignment paradigm for generative models, offering an alternative to explicit reward modeling and reinforcement learning~\cite{ouyang2022rlhf,li2025gentse,li2025aligning, wu2025mind}. DPO is a methodology derived from preference learning that reformulates alignment as supervised learning over preferred and dis-preferred sample pairs, enabling stable optimisation. Unlike traditional reinforcement-learning-based human feedback methods, which first train a separate reward model and then optimise a policy with reinforcement learning, DPO solves the alignment objective with a simple classification loss. Experimental results in recent TTS works showed that DPO not only simplifies training but also matches the performances in controlling sentiment and improving response quality~\cite{tian2025preference_alignment_tts}. 

Given a transcript $A$, and $\hat{X}^{+}$ denotes a preferred synthetic speech candidate and $\hat{X}^{-}$ denote a dis-preferred candidate under the same conditioning input. DPO maximizes the relative preference margin via:
\begin{align}
\mathcal{L}_{\mathrm{DPO}}\!
=
-\mathbb{E}\Bigg[\!
\log\!\sigma \Big(
\beta\!\log \frac{\pi_\theta(\hat{X}^{+}\!\mid\!A)}{\pi_{\mathrm{ref}}(\hat{X}^{+}\!\mid\!A)}
\!-\!\beta\!\log \frac{\pi_\theta(\hat{X}^{-}\!\mid\!A)}{\pi_{\mathrm{ref}}(\hat{X}^{-}\!\mid\!A)}
\Big)
\Bigg]
\end{align}
where $\pi_\theta$ is the trainable TTS model, 
$\pi_{\mathrm{ref}}$ is a frozen reference TTS model that stabilises optimisation, 
$\beta > 0$ is a temperature parameter controlling the strength of the preference margin, 
$\sigma(\cdot)$ denotes the sigmoid function, and 
$\mathbb{E}$ represents expectation over the preference dataset. The target TTS model $\pi_\theta$, initialised from $\pi_{\mathrm{ref}}$, receives the same transcript $A$ and computes sequence probabilities.

The DPO formulation enables flexible integration of task-specific preference signals without requiring explicit reward modeling, making it particularly suitable for extending preference learning beyond traditional natural language generation settings. While preference-based methods have achieved strong results in natural language generation and large language model alignment, their application to code-switching speech synthesis remains relatively limited. Existing work on preference-guided TTS primarily focuses on perceptual quality, naturalness, or intelligibility metrics, rather than fine-grained linguistic properties. In particular, linguistically grounded measures of code-switching, computed from synthesised speech, have not been investigated as preference objectives for acoustic-level TTS alignment, nor has their impact on downstream code-switching ASR been evaluated.

\section{Proposed Methodology}

\subsection{Acoustic-level CMI Speech}
In this paper, our proposed framework follows the language alignment strategy proposed in \cite{liu2025language_alignment}, 
generating pseudo frame-level language labels directly from the decoder cross-attention without requiring forced alignment or manual annotations. Specifically, averaged cross-attention from the last decoder layer provides a frame-to-token alignment, allowing token-level language identities to be predicted onto encoder frames as pseudo labels. These labels enable acoustic frame-level Language Identification (LID) and and is done through training with the Language Alignment Loss (LAL) proposed by the paper.

With the capability to pseudo-label acoustic frames of synthetic speech, this paper extends CMI to the speech domain to better measure preservation of code-switching patterns in synthetic speech. CMI measures the proportion of tokens that do not belong to the dominant language in a text sequence. Inspired by this definition, we extend CMI to the speech domain by treating each acoustic frame as a language-labelled unit, as
\begin{align}
\mathrm{CMI}_{\text{speech}}(u)
=
\frac{T(u) - \max_{k \in \mathcal{L}} T_k(u)}{T(u)},
\label{eq:cmi_speech_u}
\end{align}
where \(T(u)\) denotes the total number of acoustic frames in utterance \(u\), and \(T_k(u)\) represents the number of frames in \(u\) assigned to language. \(k \in \mathcal{L}\), such that \(T(u)=\sum_{k \in \mathcal{L}} T_k(u)\). \(\mathcal{L}\) denotes the set of language labels considered in the utterance. A higher \(\mathrm{CMI}_{\text{speech}}(u)\) indicates higher distribution of languages across frames, reflecting stronger acoustic code-mixing within the utterance. Unlike the traditional token-level CMI computed over discrete textual units, \(\mathrm{CMI}_{\text{speech}}\) is defined at the acoustic frame level, enabling direct measurement of acoustic language synthesis and fine-grained frame-level language transitions in speech.

After which, we measure the \(\Delta_{\text{CMI}}\) to quantify the preservation of code-switching characteristics. We compute this metric for between ground-truth and synthetic speech. The \(\Delta_{\text{CMI}}\) is defined as
\begin{align}
\Delta_{\text{CMI}}
=
\left|
\mathrm{CMI}_{\text{speech}}(\hat{X})
-
\mathrm{CMI}_{\text{speech}}(y)
\right|.
\end{align}
\(\hat{X}\) denotes the synthetic speech signal generated from the transcript, and \(y\) denotes the corresponding ground-truth speech signal. \(\Delta_{\text{CMI}}\) compares their frame-level code-mixing structure via \(\mathrm{CMI}_{\text{speech}}\).
Lower values indicate better preservation of overall code-switching proportion between the synthesised speech and ground truth speech.

\subsection{Pipeline}
After deriving \(\Delta_{\text{CMI}}\), this paper proposes a DPO framework to explicitly align synthetic code-switching speech with linguistic structure and perceptual quality. A multi-critic DPO framework is introduced during TTS training. As illustrated in Fig.~\ref{fig:pipeline}, the optimised TTS model is ultimately used to generate synthetic code-switching speech for data augmentation. The resulting speech is then incorporated into downstream ASR training, improving recognition robustness and reducing cross-language confusion. To obtain this optimised TTS, the paper proposes the pipeline to consist of 3 stages, firstly, basic fine-tuning to adapt the reference TTS model to the CS task. Secondly, DPO alignment, which will be further described in the further sections. Third, ASR training where the resultant optimised TTS model's synthetic speech will then be combined with original ground truth training data to fine-tune the CS ASR system. The proposed CS DPO framework consists of the following:

\begin{figure}[t]
  \centering
  \includegraphics[width=\linewidth]{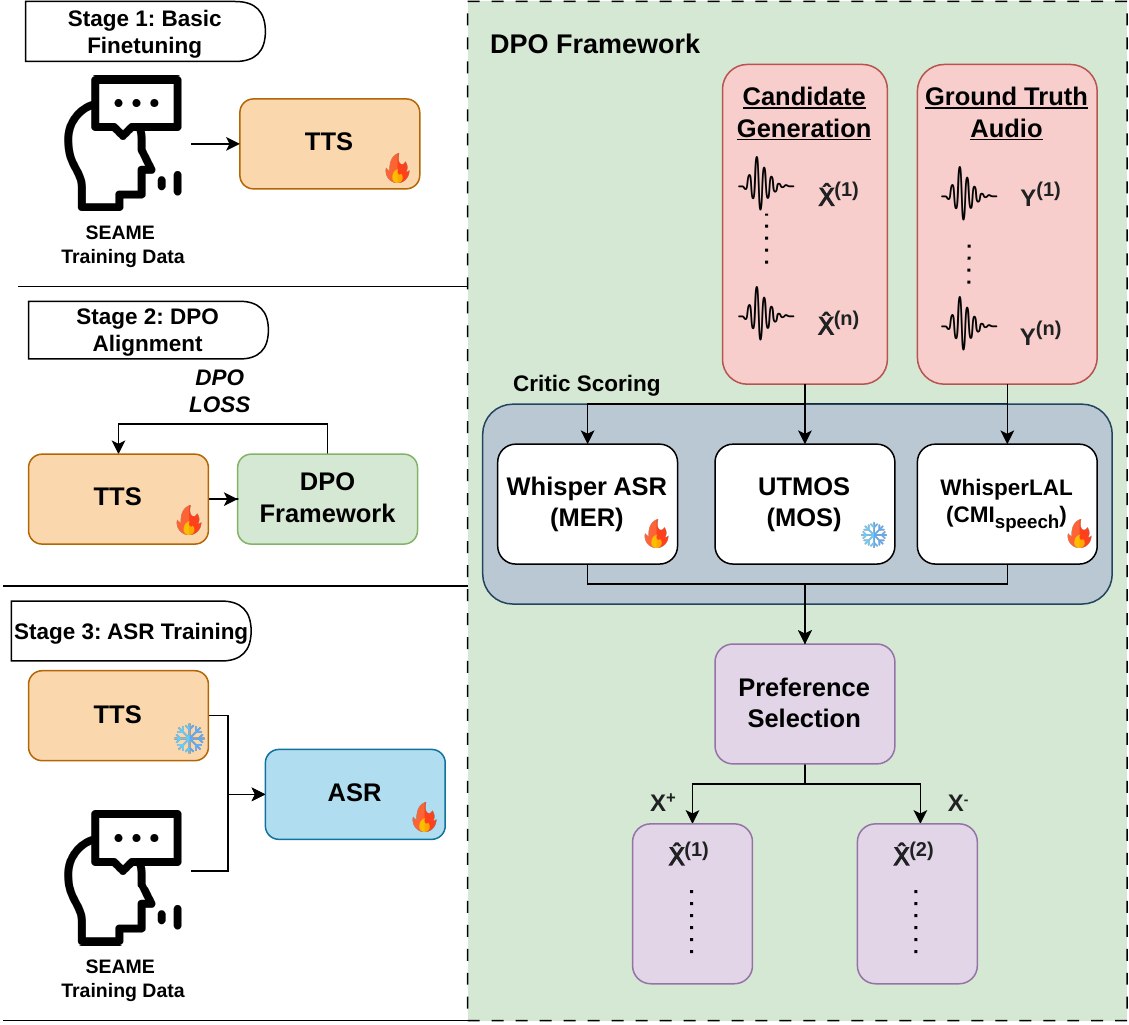}
  \caption{Overview of the proposed DPO-based TTS alignment framework.}
  \label{fig:pipeline}
\end{figure}

\subsubsection{Candidate Generation}
For each transcript \(A\), the fine-tuned TTS model \(\pi_{\mathrm{ref}}\) generates \(N\) synthetic speech candidates \(\{\hat{X}^{(n)}\}_{n=1}^{N}\) via stochastic sampling. Candidates share the same transcript and prompt speech, but differ in acoustic realization and prosody due to sampling randomness. The temperature modulates the sampling distribution and controls the degree of diversity.

\subsubsection{Critic Scoring}
After generation, each candidate $\hat{X}^{(n)}$ is evaluated using automatic critics to capture intelligibility, perceptual quality, and code-switching fidelity. First, the framework computes the mixed error rate (MER) using an ASR model by comparing the transcription of $\hat{X}^{(n)}$ with the reference transcript, where lower MER indicates downstream ASR compatibility. Second, perceptual naturalness is estimated using a pretrained UTMOS~\cite{saeki2022utmos} predictor, which provides a non-intrusive speech quality score without requiring human annotation, with higher values reflecting better naturalness. 

Finally, this paper proposes the \(\mathrm{CMI}_{\text{speech}}\) as an explicit critic for code-switching fidelity. Specifically, we measure the discrepancy \(\Delta_{\text{CMI}}\) between the synthetic and ground-truth utterances, and prefer candidates with smaller \(\Delta_{\text{CMI}}\). This method integrates an acoustic, frame-level code-mixing metric into preference learning for code-switched TTS, directly guiding generation toward more realistic language-boundary synthetic speech.
\subsubsection{Preference Selection}

To combine heterogeneous critic signals, all metrics are first converted to a unified format through normalization with scores ranging from 0 to 1 for each criteria, the final ranking score is defined as
\begin{align}
R(\hat{X}) =
\lambda\tilde{S}_{\text{UTMOS}}(\hat{X})
-
\gamma\tilde{S}_{\text{MER}}(\hat{X})
-
\nu\tilde{S}_{\Delta\text{CMI}}(\hat{X})
\end{align}
where \(\tilde{S}_{\text{UTMOS}}(\hat{X})\), \(\tilde{S}_{\text{MER}}(\hat{X})\), and \(\tilde{S}_{\Delta\text{CMI}}(\hat{X})\) are the normalised critic scores, and \(\lambda,\gamma,\nu \ge 0\) are their corresponding weights.

Preference pairs are constructed by ranking candidates of each transcript according to the normalized critic score. To provide a stable supervision signal, we pair the highest-ranked candidate with the lowest-ranked candidate for each transcript, forming a maximally contrasted pair for DPO optimization. To further improve stability, we apply threshold-based filtering to remove unreliable candidates. Specifically, after forming the pairs, pairs with the preferred candidate with MER greater than 20\%, UTMOS lower than 2.5, or CMI$_{\text{speech}}$ difference exceeding 20\% is discarded as they provide an unstable alignment to the training step.

\section{Experiment Setup}

The experiments in this paper are conducted on the SEAME corpus, a benchmark for conversational Mandarin-English code-switching speech recognition~\cite{liu2025csasr_lens}. SEAME contains approximately 192 hours of spontaneous conversational speech recorded from over 150 bilingual speakers in Singapore and Malaysia, where Mandarin and English are frequently mixed within and across utterances. The corpus exhibits both intra-sentential and inter-sentential code-switching under informal speaking conditions, making it particularly challenging for ASR systems. TTS training text and prompts are drawn only from SEAME, with no external text data.

For speech synthesis, we adopt CosyVoice2\cite{du2024cosyvoice2}, a multilingual LLM-based auto-regressive TTS model to incorporate DPO training. In our experiments, CosyVoice2 is first fine-tuned on the SEAME conversational Mandarin-English code-switching speech to adapt it to mixed-language conversational acoustics. Fine-tuning is performed using the AdamW optimizer~\cite{loshchilov2019adamw} with a learning rate of $2\times10^{-4}$ and a linear warm-up schedule. Training is conducted for approximately 50k steps with a batch size of 4 utterances, and early stopping is applied based on validation loss to prevent overfitting. The fine-tuned model is then used to synthesise CS speech for ASR data augmentation.

For ASR, we use Whisper ASR model as the downstream evaluation system. Whisper takes log-Mel spectrograms as input and employs a Transformer-based architecture to jointly model acoustic and linguistic information. We fine-tune the Whisper-\emph{large v3} model on SEAME conversational Mandarin--English code-switching speech, augmented with synthetic speech generated by the TTS model. Fine-tuning is carried out using the Adam  optimizer~\cite{kingma2015adam} with a learning rate of $1\times10^{-5}$. Models are trained till convergence with a batch size of 1 utterance per A40 GPU. All ASR experiments use identical training and decoding configurations across different augmentation strategies to ensure fair comparison. In particular, it is ensured that summed duration of real and synthetic speech is consistent.

\subsection{Scoring models}

For MER, a SEAME fine-tuned Whisper-\emph{largev3} ASR model is used to provide intelligibility-based preference signals for DPO. For each pair of synthesised speech samples generated from the same transcript, we decode the audio using Whisper and compute MER. Preference labels are assigned by favouring the sample with lower MER, encouraging the TTS model to generate speech that is more accurately recognized by a strong downstream ASR system.

For UTMOS score, UTMOS as an automatic speech quality predictor is used to provide perceptual preference signals for DPO. For each pair of synthesised samples, UTMOS assigns a predicted MOS score, and the sample with higher predicted quality is preferred. These preferences guide the TTS model towards generating more natural and perceptually pleasing speech without requiring human listening tests.

For \(\Delta_{\text{CMI}}\) calculation, to encourage realistic code-switching behavior during training, a SEAME based fine-tuned Whisper model with LAL loss is used to generate pseudo frame-level language labels from synthesised speech. Specifically, Whisper provides language identification predictions that are aligned to acoustic frames, enabling automatic identification of contiguous language regions within the waveform. These pseudo labels are then used to compute $\mathrm{CMI}_{\text{speech}}$, which quantifies the realised degree of language mixing directly from the synthesised audio. 

\section{Results}

\begin{table}[t]
  \caption{Cosyvoice TTS Performance Comparison Across DPO Optimization Metrics (DevMAN and DevSGE test set)}
  \label{tab:dpo_ablation_transposed}
  \centering
  \begin{tabular}{l l | c c c}
    \toprule
    {Model} & {Reward} 
    & {UTMOS} $\uparrow$ 
    & {MER} $\downarrow$ 
    & \(\Delta_{\text{CMI}}\) $\downarrow$ \\
    \midrule
    CosyVoice & -- & 3.1 & 16.2 & 28.1 \\
    \midrule
    DPO & MER & 3.2 & 14.9 & 25.7 \\
        & + UTMOS & 3.8 & 13.2 & 21.9 \\
        & \hspace{1.1mm} + \(\Delta_{\text{CMI}}\) 
        & \textbf{3.8} & \textbf{10.3} & \textbf{16.1} \\
    \bottomrule
  \end{tabular}
\end{table}

Table~\ref{tab:dpo_ablation_transposed} shows how progressively adding critic signals to DPO better aligns CS TTS with our objective of generating better quality synthetic speech that effectively improves downstream CS-ASR. The table shows the results after DPO fine-tuning CosyVoice TTS on SEAME training set and reproducing DevMAN and DevSGE test sets with the various different reward critics. Optimising with MER alone improves intelligibility (16.2\% → 14.9\%) but only modestly reduces \(\Delta_{\text{CMI}}\) (28.1\% → 25.7\%). Adding UTMOS further improves perceptual quality (3.1 → 3.8) and reduces MER to 13.2\%, indicating that DPO alignment enhances overall synthesis quality beyond ASR compatibility.

Importantly, incorporating the proposed acoustic-level \(\Delta_{\text{CMI}}\) critic yields the largest reduction in code-switch difference (28.1\% → 16.1\%) while simultaneously achieving the lowest MER (10.3\%) without degrading perceptual quality. This demonstrates that explicitly aligning acoustically-perceived language proportion improves structural faithfulness of code-switching speech. Since our objective is to assess how well the TTS system preserves intelligibility, naturalness, and code-switch structure, the consistent improvements across all three metrics confirm the effectiveness of the proposed DPO framework in improving the CS TTS quality.

\begin{table}[th]
  \caption{Mixed Error Rate (MER, \%) on SEAME DevMAN and DevSGE test sets.
   Real: Ground truth SEAME training data. For Cosyvoice and Cosyvoice with DPO: 200 hours (100 hours of Real + 100 hours of Synthetic Audio Data)
  }
  \label{tab:seame_transposed_clean}
  \centering
  \begin{tabular}{l|c c}
    \toprule
    {Train Configuration} & {DevMAN} & {DevSGE} \\
    \midrule
    \multicolumn{3}{l}{{Whisper ASR}} \\
    \midrule
    \textit{Real (100h)} & 12.1 & 17.8 \\
    \textit{+ CosyVoice} & 10.1 & 16.0 \\
    \hspace{1em}\textit{+ DPO (UTMOS, MER)} & 9.6 & 15.1 \\
    \hspace{1em}\textit{+ DPO (UTMOS, MER, \(\Delta_{\text{CMI}}\))} 
        & \textbf{8.9} & \textbf{14.2} \\
    \midrule
    \multicolumn{3}{l}{{CTC-Based Conformer}} \\
    \midrule
    \textit{Real (100h)} & 16.8 & 23.6 \\
    \textit{+ CosyVoice} & 16.1 & 22.8 \\
    \hspace{1em}\textit{+ DPO (UTMOS, MER)} & 15.8 & 22.3 \\
    \hspace{1em}\textit{+ DPO (UTMOS, MER, \(\Delta_{\text{CMI}}\))} 
        & \textbf{15.4} & \textbf{21.9} \\
    \bottomrule
  \end{tabular}
\end{table}
Table~\ref{tab:seame_transposed_clean} reports MER on SEAME DevMAN and DevSGE for both Whisper ASR and a ESPNet's CTC-based Conformer~\cite{watanabe18_interspeech} under different data augmentation strategies to show effects of downstream ASR training. Adding an equal proportion of 100 hours of synthetic speech generated from CosyVoice fine-tuning consistently reduces MER compared to the 100-hour real-data baseline (12.1\%/17.8\% → 10.1\%/16.0\% for Whisper; 16.8\%/23.6\% → 16.1\%/22.8\% for Conformer), demonstrating that high-quality synthetic data can improve ASR robustness for conversational code-switching speech.

Applying DPO yields consistent further improvements across both evaluation sets and architectures, suggesting that preference-optimised synthetic speech is more effective than simple generated augmentation. Incorporating the proposed  \(\Delta_{\text{CMI}}\) critic produces the largest gains compared to traditional DPO methods of only using MER and UTMOS, reducing MER to 8.9\%/14.2\% for Whisper ASR and 15.4\%/21.9\% for the Conformer on DevMAN/DevSGE, respectively. The consistent improvements observed across models demonstrate that explicitly preserving acoustic-level code-switching structure during TTS generation substantially enhances downstream ASR performance.

\subsection{Ablation Analysis}
\begin{table}[t]
\caption{Qualitative comparison of code-switching synthesis under different training strategies by Whisper ASR.}
\label{tab:qualitative_cs}
\centering
\small
\renewcommand{\arraystretch}{0.95}
\begin{tabular}{l|p{5.2cm}}
\toprule
\textbf{Method} & \textbf{Output} \\
\midrule

Ground Truth &
\begin{CJK}{UTF8}{gbsn}
他 spend too long spend too long 去 做
\end{CJK}
\newline
\begin{CJK}{UTF8}{gbsn}
o hall olympiad 是 另 外 一 个 也 是
\end{CJK}
\newline
\begin{CJK}{UTF8}{gbsn}
什 么 science 的 but 他 有 一 点 
\end{CJK} \\
\midrule

CosyVoice2 FT &
\begin{CJK}{UTF8}{gbsn}
他 spend two 龙 spend two 龙 去 做
\end{CJK}
\newline
\begin{CJK}{UTF8}{gbsn}
oh hall 奥 林 匹 克 是 另 外 一 个 也 是
\end{CJK}
\begin{CJK}{UTF8}{gbsn}
什 math science the but 他 有 一 点 
\end{CJK} \\
\midrule

\begin{tabular}[t]{@{}l@{}}
CosyVoice2 FT \\
+ DPO \\
(MER, UTMOS)
\end{tabular}
&
\begin{CJK}{UTF8}{gbsn}
他 spend too 龙 spend too 龙 去 做
\end{CJK}
\newline
\begin{CJK}{UTF8}{gbsn}
o hall 奥 林 匹 克 是 另 外 一 个 也 是
\end{CJK}
\begin{CJK}{UTF8}{gbsn}
什 math science 的 but 他 有 一 点 
\end{CJK} \\
\midrule

\begin{tabular}[t]{@{}l@{}}
CosyVoice2 FT \\
+ DPO \\
(MER, 
UTMOS, \\
\(\Delta_{\text{CMI}}\))
\end{tabular}
&
\begin{CJK}{UTF8}{gbsn}
他 spend too long spend too long 去 做
\end{CJK}
\newline
\begin{CJK}{UTF8}{gbsn}
o hall olympiad 是 另 外 一 个 也 是
\end{CJK}
\newline
\begin{CJK}{UTF8}{gbsn}
什 么 science 的 but 他 有 一 点 
\end{CJK} \\
\bottomrule
\end{tabular}
\end{table}
To futher evaluate on the effects of \(\Delta_{\text{CMI}}\), Table~\ref{tab:qualitative_cs} shows the progressive improvement based on the different methods that has been used to improve the TTS. Basic fine-tuning still has CS errors, where English words are similar shifted in Mandarin and language boundaries become unstable. While DPO with MER and MOS improves overall intelligibility, cross-language substitutions remain, indicating that recognition- and perceptual-based rewards alone do not sufficiently constrain code-switching structure. By incorporating \(\Delta_{\text{CMI}}\), the model guides the language distribution in synthesised audio, preserving intended language boundaries and restoring correct pronunciation of English or Chinese CS segments. This demonstrates that $\mathrm{CMI}_{\text{speech}}$ guided preference optimization effectively produces more accurate CS speech.

\section{Conclusion}
In this work, we presented a CS metric guided DPO framework for improving CS TTS. Our approach explicitly aligns synthetic speech with intelligibility, perceptual quality, and realistic language-mixing structure. By integrating \(\Delta_{\text{CMI}}\) for measuring synthesised CS complexity, we construct contrastive preference pairs through normalized scoring and threshold-based filtering. Experimental results on SEAME demonstrate consistent improvements across both TTS and downstream ASR evaluation, including substantial reductions in MER while maintaining higher perceptual quality and better preservation of language mixing. These findings highlight the effectiveness of multi-critic preference learning in enforcing structured CS behavior and suggest that CS alignment-driven optimisation provides a promising direction for CS TTS generation to improve CS ASR.

\section{Generative AI Use Disclosure}
Generative AI tools were used for limited assistance with manuscript editing and presentation (e.g., grammatical validation, removal of redundant sentences and phrases, preparing LaTeX equations and LaTeX formatting suggestions). The literature review and all scientific contributions, including but not limited to problem formulation, methodology, experiments, results, and conclusions, were performed by the authors. All authors reviewed the manuscript and are responsible for the final submission.

\bibliographystyle{IEEEtran}
\bibliography{mybib}

\end{document}